\pdfoutput=1

\documentclass[aps,prd,amsmath,notitlepage,superscriptaddress,%
              nobibnotes,nofootinbib,preprintnumbers,%
              onecolumn,12pt,tightenlines]{article}

\usepackage{graphicx}

\usepackage{verbatim}

\graphicspath{{figures/}}

\usepackage[normalem]{ulem}
\usepackage[usenames]{color}
\definecolor{DarkRed}{rgb}{0.5,0.0,0.0}
\definecolor{DarkGreen}{rgb}{0.0,0.5,0.0}
\definecolor{DarkBlue}{rgb}{0.0,0.0,0.5}
\definecolor{Magenta}{rgb}{1.0,0.0,1.0}
\definecolor{DarkMagenta}{rgb}{0.5,0.0,0.5}
\definecolor{Orange}{rgb}{1.0,0.5,0.0}
\definecolor{DarkOrange}{rgb}{0.8,0.3,0.0}
\definecolor{DarkCyan}{cmyk}{1.0,0.0,0.0,0.5}
\definecolor{Brown}{cmyk}{0.0,0.8,1,0.6}

\usepackage{soul}

\usepackage[colorinlistoftodos,textwidth=2in]{todonotes}
\usepackage{jcappub}

\newcommand{\refapp}[2][sec:]{the Appendix}
\newcommand{\Refapp}[2][sec:]{The Appendix}

\usepackage{amsmath,amssymb,amsfonts,amsthm}
\usepackage{hyperref}

\usepackage[sorting=none]{biblatex}
\bibliography{mDMbib.bib}

\usepackage{sectsty}
\sectionfont{\centering}

\notoc
\begin{document}


\title{DAMA annual modulation is not due to electron recoils from Plasma/Mirror Dark Matter with Kinetic Mixing}

\author[a]{Douglas Q. Adams}

\author[b]{Sunniva Jacobsen}

\author[c]{Chris Kelso,}

\affiliation[a]{Department of Physics,
University of South Carolina,
Columbia, SC 29208}
\affiliation[b]{
    Oskar Klein Centre for Cosmoparticle Physics,
Physics Department,
Stockholm University,
SE-106 91 Stockholm, Sweden}
\affiliation[c]{Department of Physics,
  University of North Florida,
  Jacksonville, FL 32224}

\date{\today}
 




\abstract{ 

If dark matter is composed of new fundamental particles, 
Earth's orbital motion around the Sun may induce an annual modulation in the rate at which these particles interact in a terrestrial detector.  
The DAMA collaboration has identified at a 12$\sigma$ confidence level such an annual modulation in their event rate. 
One previously proposed explanation is `plasma dark matter' (e.g. mirror dark matter) 
which would have electron recoils in the detector from interactions via kinetic mixing.
We perform a chi-squared goodness of fit test of this plasma dark matter model to DAMA/LIBRA  modulation amplitude which rejects the hypothesis of plasma dark matter as an explanation of this data at greater than 7$\sigma$.
} 

\maketitle


\section{Introduction}
\label{sec:Intro}
The nature of dark matter continues as an extremely intriguing unsolved mystery in modern physics.  One of the most exciting explanations is that dark matter is a new fundamental particle and there is currently an extensive experimental effort to try to detect the particle nature of dark matter (DM) through its interactions with Standard Model particles \cite{Drees:2012ji,Beringer:1900zz}.  
More than thirty years ago, a mechanism was proposed for detecting neutrinos ~\cite{Drukier:1983gj} and Weakly Interacting Massive Particles (WIMPs) \cite{Goodman:1984dc} via coherent elastic scattering with nuclei.  Soon after, in Ref.~\cite{Drukier:1986tm}, detection rates in the context of a Galactic Halo of WIMPs were first computed and it was proposed that DM particles could be differentiated from background by looking for an annual modulation of the signal.

Since 1995 the DAMA (and subsequently DAMA/LIBRA) collaboration has been taking data with a  ``direct detection'' experiment to search for the annual modulation of nuclear recoils due to dark matter scattering within NaI scintillation detectors.  DAMA has identified at a 12$\sigma$ confidence level an annual modulation in their event rate that is consistent with a dark matter signal.  Thus far, no plausible alternative explanation for the modulation signal observed in the DAMA detectors has been universally accepted~\cite{Bernabei:2012wp}.  Many in the scientific community remain skeptical of the dark matter interpretation of the DAMA annual modulation signal due to the very stringent null results from numerous other detectors (see Ref.~\cite{Schumann:2019eaa} for a recent review).     However, since the other experiments are not made of the same material as DAMA, it remains crucial for new experiments also consisting of NaI crystals to
see whether or not they are able to reproduce the DAMA's signal.

This has inspired are several upcoming NaI experiments that will test the DAMA modulation results. SABRE~\cite{Antonello_2019}, which will consist of two identical experiments located in the northern and southern hemispheres and will feature active background rejection. The ANAIS-112 experiment has been taking data since August of 2017 and already released null results for a modulation signal when analyzing 1.5 years of data~\cite{Amare:2019jul}.  The experiment is on schedule to be sensitive at the $3\sigma$ level to the amplitude of DAMA's modulation after 5 years of data taking. The  COSINE-100\cite{ Ha:2017egw}  is a joint effort between the DM-Ice~\cite{deSouza:2016fxg} and KIMS~\cite{Kim:2012rza} collaborations, featuring a active liquid scintillator veto has been running since September 2016 and also released their first results in a search for a modulation signal analyzing 1.7 years of data~\cite{Adhikari:2019off}.  Their results are still consistent with zero modulation, although there is some very slight preference (that is not statistically significant) for a modulation of similar amplitude and phase to that of the DAMA signal.  Similar to ANAIS-112, COSINE-100 anticipates sensitivity to the DAMA modulation amplitude at the $3\sigma$ level after 5 years of data taking.  Another exciting prospect is the COSINUS experiment that will use cryogenically cooled NaI so that the detector will be able to read a phonon/heat signal in addition to measuring scintillation light \cite{Reindl:2017bun}.  

Previously Foot proposed an explanation for the DAMA signal in terms of ``plasma dark matter"~\cite{Foot:2014uba, Clarke:2015gqw} causing electron rather than nuclear recoils in the detector. 
In this model, there is a dark sector in the Universe in parallel to our Standard Model (SM) sector. The dark matter of the Milky Way consists of a dark plasma of dark electrons and dark protons (much heavier than the dark electrons) coupled together via a massless dark photon ~\cite{Foot:2014mia}.
The dark electrons interacting with electrons in the detector could produce electron recoils with the observed keV energies~\cite{Clarke:2015gqw, Foot:2014xwa}. Assuming energy equipartition of this two component system implies large velocities of the dark electrons in excess of the escape velocity from the Galaxy (since they are much lighter than the dark protons), with the particles bound to the halo mainly by the dark electromagnetic force rather than by gravity.  Such high velocities are required to satisfy the necessary kinematics to produce detectable electron recoils in DAMA (note that such rapid velocities are not possible in the Galaxy for Weakly Interacting Massive Particles).

The plasma model assumes that DM arises from a hidden sector that interacts with SM particles only via gravity and kinetic mixing of the SM photon and hidden photon, with a Lagrangian
\begin{equation}
L = L_{SM} + L_{dark} + \frac{\epsilon}{2} F^{\mu\nu} F'_{\mu\nu} 
\end{equation}
where $L_{dark}$ is the Lagrangian for the dark sector containing dark electrons, dark protons, dark photons, and other dark particles in parallel with the SM.   Here $F^{\mu\nu} [F'_{\mu\nu}]$ is the U(1) [U'(1)] field strength tensor of the photon [hidden photon] and $\epsilon$ is a dimensionless free parameter that has been shown in previous work to be very small in this model, $\epsilon = 10^{-9} - 10^{-10}$~\cite{Foot:2018DAMA}. Due to the kinetic mixing, the dark electrons obtain a small SM electric charge, $\epsilon$e.

A particular example of this model is `mirror dark matter,' where the dark particles have the same masses as their SM counterparts, e.g. $m_e' = m_e$ etc, where prime refers to particles in the dark sector ~\cite{Foot:2014mia, Foot:1991bp}.  

Calculation of expected direct detection rates in this model is complicated by the interaction between the usual DM halo wind and a `dark sphere' of captured dark matter in the Earth, which strongly affects the dark matter distribution entering the detector. Estimates were made in previous work of the resulting DM velocity distribution~\cite{Clarke:2015gqw,Foot:2014mia,  Foot:2018DAMA}. In this scenario, one would naively expect the temperature of the dark halo to have $T\approx 0.5\,$keV, which would make producing the DAMA signal in the 1--6keV energy range difficult.  However, ``dark interactions" between the halo DM and the DM bound to the earth could lead to large distortions of the velocity distribution of the mirror electrons that might enhance these larger energy collisions as discussed in Ref.\cite{Foot:2018DAMA}. Additionally, if the exact mirror symmetry is removed and more generic hidden sector models are considered, the dark electron temperature can easily be above 6\,keV.  In particular, if the dark proton has a mass above $\sim20$\,GeV, this condition is satisfied. Also, increasing the mass of the dark electron above that of the standard model electron would increase the energy of scatters in the detector.  Our analysis is general in the sense that it does not require an assumption of any specific choice of masses of the dark sector particles.

Previously Foot suggested that the plasma dark matter model could be consistent with both the DAMA modulation data and the constraints from other experiments, including XENON100 and DarkSide-50~\cite{Foot:2018DAMA} .  The LUX experiment placed constraints on the mirror matter plasma model \cite{Akerib:2019diq}.
More recently, another paper \cite{Zu:2020bsx} argued in favor of explaining the recent XENON1T excess of low energy electron recoil events in an S1 analysis near their $\sim 2$keV threhold \cite{Aprile:2020tmw}.  Simultaneous with our current paper, Foot \cite{new} now finds that the model is ruled out by XENON S2 data.

\section{Expected Signal}
Because the dark electrons are electrically charged, they can scatter off ordinary electrons in the detector.
If we approximate the target electron as free and at rest relative to the incoming dark electron of speed v, 
then the scattering rate of $e' - e$ scattering can be described as a Coulomb process:
\begin{equation}
    \frac{d\sigma}{d E_R} = \frac{g_T(E_R) \lambda}{E_R^2 v^2} \ ,
\end{equation}
where $\lambda = 2\pi\epsilon^2\alpha^2/m_e$, $v$ is the relative velocity between the electrons, $E_R$ is the recoil energy, $g_T(E_R)$ is the number of free electrons in target $T$ (i.e. with binding energy smaller than $E_R$),  and $\alpha$ is the fine-structure constant. 
The rate for $e-e'$ scattering then becomes,
\begin{equation}
\begin{aligned}
    \frac{d R_e}{d E_R} 
    = g_T(E_R) N_T n_{e'}\frac{\lambda}{E_R^2}I(\vec{v}_e, \theta) \ ,
\end{aligned}
\end{equation}
where $n_{e'}$ is the mirror electron number density at the detector location, $N_T$ is the number of target atoms per detector mass, $\vec{v}_E$ is the velocity of the Earth around the Sun and the velocity integral is
\begin{equation}
\label{eq:velocityintegral}
    I(\vec{v}_E, \theta) = \int_{v>v_{\rm min}}^\infty \frac{f(\vec{v};\vec{v}_E;\theta)}{|\vec{v}|}d^3 v \ .
\end{equation}
The minimum velocity required for scattering to occur is  $v_{\rm min} = \sqrt{2E_R/m_e}$.
As described above, the velocity distribution $f(\vec{v};\vec{v}_E;\theta)$ of dark electrons which arrive at the detector is nonstandard as it depends in detail on the ``dark ionosphere"  and has been 
described in Ref.\cite{Foot:2018DAMA} in references therein. 
Further, the velocity distribution varies throughout the day and year due to the Earth's daily rotation and its motion around the Sun. This gives rise to diurnal and annual modulation ~\cite{Freese:2012xd} of the expected dark matter signal in the detector. 
 
When the mean velocity of the dark matter particles is much larger than the minimum velocity required for scattering, we can approximate the velocity integral as independent of the recoil energy. Since $v_{\rm min} \propto \sqrt{E_R}$, this assumption only holds when $E_R < E_R^T$, where $E_R^T$ is some threshold energy. 
Taking $E_R^T \geq 6$keV (the largest recoil energy in the DAMA data), this limit requires $m_p'  > 20$ GeV, unless the dark electron velocity distribution is significantly enhanced at higher energies as discussed above. The signal becomes strongly suppressed when $E_R > E_R^T$, falling off a much faster rate than $1/E_R^2$.
 
In the limit $E_R < E_R^T$, one can Taylor expand the velocity integral to obtain the following expression for the scattering rate:
\begin{equation}
\label{eq:countrate}
    \frac{d R_{e}}{d E_{R}}=g_{T} N_{T} n_{e^{\prime}} \frac{\lambda}{v_{c}^{0} E_{R}^{2}}\left[1+A_{v} \cos \left(\omega t-\omega t_{0}\right)+A_{\theta}(\theta-\bar{\theta})\right] \quad \text { for } E_{R} \lesssim E_{R}^{T} \ ,
\end{equation} 
where $v_c^0\equiv 1/I_0,$ the first term in the Taylor expansion of the velocity integral in Eq.(\ref{eq:velocityintegral}).
The first term in brackets is the average count rate, the second term is the annual modulation, and the third term is the diurnal modulation (not studied in this paper). 
Here $\omega = 2 \pi$/year, $A_v$ is the annual modulation amplitude, and $t_0=153$ days (June 2nd).

We will write the annual modulation of the signal in this phenomenological model for the count rate as:
\begin{equation}
\label{eq: diff recoil rate}
\frac{d R_{e}^{\rm mod}}{d E_{R}}= \bigg(  \frac{2\pi \alpha^2 g_{T} N_{T} }{ m_e } \bigg) \frac{\phi_{e^{\prime}}}{E_{R}^{2}} \cos \left(\omega t-\omega t_{0}\right) 
 \end{equation}
where the constant $\phi_{e^{\prime}}=\frac{n_{e^{\prime}}A_{v}\epsilon^2}{v_{c}^{0}}$ is introduced to contain all the dependence on the dark electron properties in Eq.~\ref{eq:countrate}. Note that since $\phi_{e^{\prime}}$ only depends on the dark electron properties, it is the same for all experiments.  As the DAMA experiment uses NaI as the detection material, we have used $g_T=56$ for the number of electrons with binding energies less than $ \sim 0.8 \,\mathrm{keV}_{\mathrm{ee}} $.  
The search for the best fit for the DAMA annual modulation data thus requires only a one parameter scan $\left(\phi_{e^{\prime}}\right)$ as described in the next section.


\section{Analysis Strategy}
We have tested how well the Foot model as described in Eq.\eqref{eq: diff recoil rate} fits with the annual modulation detected in the DAMA/LIBRA experiment~\cite{Bernabei:2018jrt}. We have performed a goodness of fit test of the modulation amplitude predicted by the Foot model with the amplitude detected by DAMA/LIBRA.

To find the predicted annual modulation signal as it would appear in the DAMA/LIBRA experiment, we need to take into account the finite energy resolution of the detector. This is done by convolving Eq.~\eqref{eq: diff recoil rate} with a Gaussian distribution:

\begin{equation}
\label{eq:dRdEee}
    \frac{d R_e^{\rm mod}}{d E_{m}} =\int_{0}^{\infty} \frac{1}{\sqrt{2\pi} \sigma_{\rm E}} e^{-( E_{m} - E_{\rm R})^2/2\sigma_E^2} \frac{d R_e^{\rm mod}}{d E_{\rm R}} d E_{\rm R}\ ,
\end{equation}
where $E_{m}$ is the recoil energy as measured by the detector and the $\sigma_{\rm E}$ is the energy resolution of the DAMA detector.   We take the energy resolution to be: $\sigma_{E}(E_m) = \alpha \sqrt{E_m} + \beta E_m$ with $\alpha=(0.448 \pm 0.035) \sqrt{\mathrm{keV}_{\mathrm{ee}}}$ and $\beta =(9.1 \pm 5.1) \times 10^{-3}\,\mathrm{keV}_{\mathrm{ee}}$ as given in Ref.~\cite{Bernabei:2008yh}.

The theoretical prediction for the average modulation amplitude in energy bin $k$ can be found by
\begin{equation}
    S_{m,k}^{th} = \sum_T \frac{\xi_T}{2(E_2 - E_1)}\int_{E_1}^{E_2} dE_m \left[ \frac{dR_e^{\rm mod}}{dE_m}(E_m, t_0) - \frac{dR_e^{\rm mod}}{dE_m}(E_m, t_0+0.5yrs) \right] \ ,
\end{equation}
where $T$ represents the atoms in the detector material, $\xi_T$ is the mass fraction of the atoms in the detector, $E_2$ and $E_1$ are the upper and lower energy limits of the bin and $t_0$ is the date when the modulation amplitude is maximal.  As a note, for the NaI detector, the ``atom" in the detector is actually a NaI molecule.

We have minimized the following $\chi^2$ function to determine the value of $\phi_{e^{\prime}}$ that gives the best fit to the DAMA modulation signal,
\begin{equation}
    \chi^2 = \sum_{k=1}^{10}\frac{(S^{th}_{m,k}(\phi_{e^{\prime}}) - S_{m,k}^{exp})^2}{\sigma_k^2} \ .
\end{equation}
Here,  $S_{m,k}^{exp}$ is the measured amplitude in DAMA in bin $k$ and $\sigma_k$ is the corresponding uncertainty in that bin. We used an alternative binning of the phase-2 DAMA data, according to the procedure in Ref.~\cite{Kelso:2013gda}. DAMA uses 36 bins, each with a width of $0.5\,$keV. The rebinning procedure groups together adjacent bins with widths that are substantially narrower than the energy resolution of the DAMA detector, and groups together all higher energy bins into one single bin. This will serve to improve the ``signal-to-noise" of the goodness of fit test that we will employ.  A detailed discussion of this rebinning procedure is presented in Ref.~\cite{Kelso:2013gda}.  The resulting bins and measured values are presented in Table \ref{tab:rebinned modulation amplitudes}.

Once the best fit point was found by minimizing the $\chi^2$ function above, we performed a $\chi^2$ goodness of fit test for this scenario. This is a hypothesis test where the test statistic is  the best fit $\chi^2$ value and the test computes the probability of  measuring a best-fit $\chi^2$ at least as large as this through random chance, assuming the best fit parameters for the model are the ``true" model.  The test statistic will follow a $\chi^2$ distribution with the number of degrees of freedom equal to the number of bins minus the number of fit parameters (9 for our case).  
\begin{table}[h!]
    \centering
    \begin{tabular}{c|c} \hline\hline
        Energy & Average $S_m$ \\ 
        $\rm [keV_{ee}]$ & $[\mathrm{cpd} / \mathrm{kg} / \mathrm{keVee}]$ \\ \hline
        $1.0-1.5$ &   $0.0232 \pm 0.0052$ \\
        $1.5-2.0$ &  $0.0164 \pm 0.0043$\\
        $2.0-2.5$ &  $0.0178 \pm 0.0028$ \\ 
        $2.5-3.0$ & $0.0190 \pm 0.0029$  \\
        $3.0-3.5$ &  $0.0178 \pm 0.0028$\\ 
        $3.5-4.0$ &  $0.0109 \pm 0.0025$\\ 
        $4.0-5.0$ &  $0.0075 \pm 0.0015$\\ 
        $5.0-6.0$ &  $0.0066 \pm 0.0014$\\ 
        $6.0-7.0$ &  $0.0013 \pm 0.0013$\\ 
        $7.0-20.0$ &  $0.0007 \pm 0.0004$\\ \hline\hline
    \end{tabular}
    \caption{Average modulation amplitudes observed by DAMA over the given energy bins, after rebinning as in Ref. \cite{Kelso:2013gda}. We have used the modulation amplitudes for the whole data sets: DAMA/NaI, DAMA/LIBRA phase-1 and DAMA/LIBRA phase-2, as presented in Ref.~\cite{Bernabei:2018jrt}  The rebinning is performed in order to improve the signal-to-noise ratio.}
    \label{tab:rebinned modulation amplitudes}
\end{table}

\section{Results}
\begin{figure}
    \centering
    \includegraphics{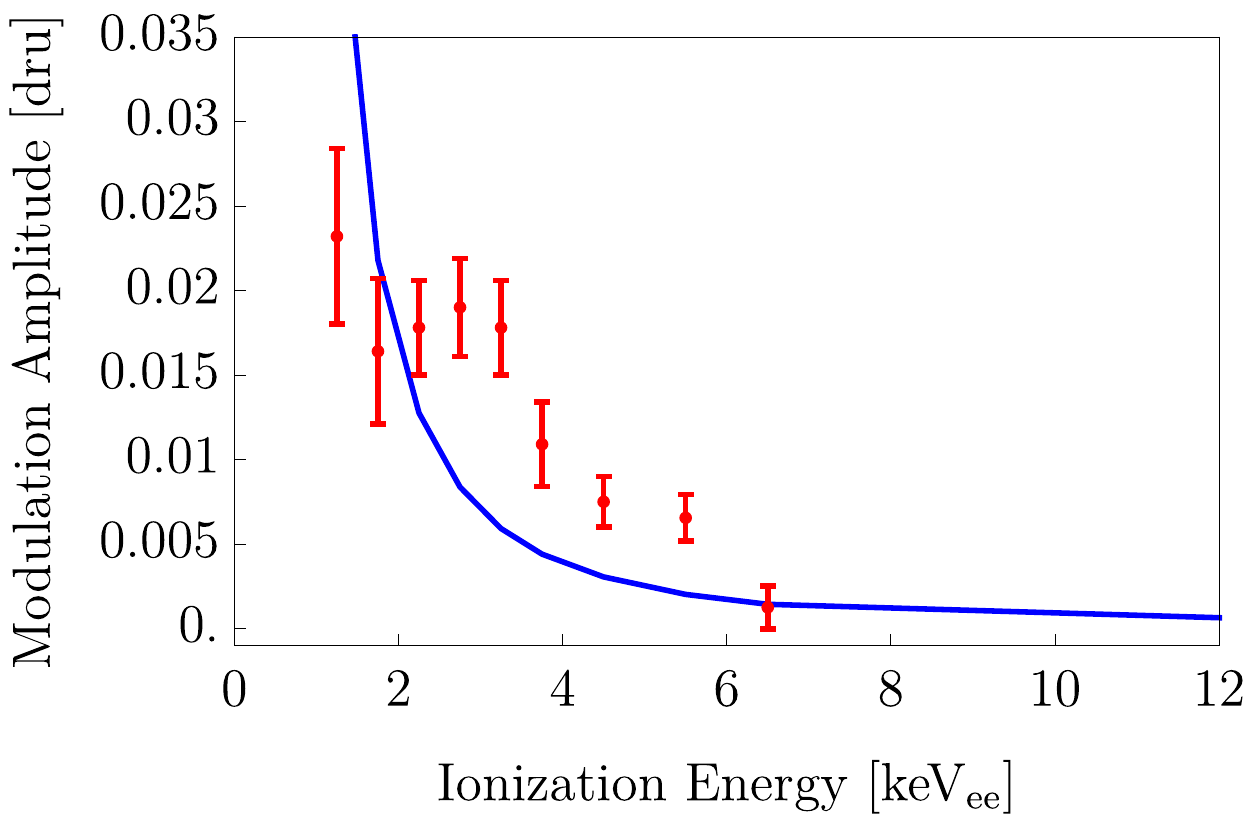}
    \caption{The predicted energy spectrum of the modulation amplitude using the best-fit Foot model (blue line) compared to the rebinned DAMA signal (red points with error bars). The best-fit Foot model is a very poor fit to the data and is excluded by the goodness-of-fit test at more than $7\sigma$.}
    \label{fig: foot model and dama signal}
\end{figure}

We find that the best fit to the DAMA data in the Foot model gives $\chi^2=81.7$ for $\phi_{e^{\prime}}=1.31\cdot10^{-32}~\rm s/cm^4$ for maximum modulation $A_v=1$.  This choice for $A_v$ would produce the fewest events in other detectors searching for the constant term of Eq.~\ref{eq:countrate} as discussed in Section~\ref{sec:OtherExp}.  The corresponding energy spectrum in the DAMA detector using the best-fit value for $\phi_{e^{\prime}}$ is presented in figure~\ref{fig: foot model and dama signal}.  Our best fit $\chi^2=81.7$ with 9 bins gives a probability of getting a $\chi^2$ at least this large equivalent to the probability of normal distribution at the $7.5\sigma$ level.  The goodness of fit test demonstrates that the Foot model is a very poor fit to the detected annual modulation in DAMA and is excluded by this test at greater the $7\sigma$ level.

 The poor fit can be traced to $1/E_R^2$ behaviour of the expected signal in the Foot model.  The DAMA signal looks like a linear decrease moving towards higher energies, while the Foot model predicts a much more rapid decay.  Additionally, the very rapid rise of the Foot model at low energies means that assuming worse resolution of the detector will exacerbate the poor fit as events from below threshold will spill over even more into the detection energy range of the detector.  This causes an even sharper rise at low energies producing an even worse fit.  On the other hand, even a DAMA-like detector with perfect energy resolution still provides a very poor fit to the data with a best fit $\chi^2=74.0$ which is still excluded at the $7\sigma$ level. 

\section{Other Experiments}
\label{sec:OtherExp}
As can be observed from Eq.~\ref{eq:countrate}, we see the annual modulation signal and the unmodulated (average) rate will both scale with the same factor of $\phi_{e^{\prime}}$, up to the $A_v$ term which will be between 0 and 1.  This means that once the Foot model is used to fit the DAMA/LIBRA data, then it will make definite predictions for signals in other direct detection experiments that are sensitive to electron recoils.

The LUX experiment searched for mirror dark matter using an exposure 95 live-days$\times$118 kg of data collected in 2013~\cite{Akerib:2019diq}.  In this work, they excluded mirror electrons with temperatures above 0.3~keV.  As noted previously, this temperature constraint can be evaded if the mirror symmetry is not exact and the dark proton and/or dark electron has a heavier mass than their standard model counterparts.  The best fit Foot model to the DAMA data of $\phi_{e^{\prime}}=1.31\cdot10^{-32}~\rm s/cm^4$ can then be used to predict the expected signal in this LUX data set due to the non-modulating (constant) term in Eq.~\ref{eq:countrate}.  The Foot model with $g_T= 45$ for the number of electrons with binding energies below $\sim 0.3\,\mathrm{keV}_{\mathrm{ee}}$ for xenon would predict approximately 150 events in the energy range of 2--4~$\mathrm{keV}_{\mathrm{ee}}$ using $A_v=1$.  Again, any smaller value for $A_v$ would scale up the number of expected events in the detector by 1 divided by the new choice for $A_v$.  The experiment measured $\sim$100 events total in this energy window, which includes all backgrounds.  This indicates that the best fit model to the DAMA data is in strong tension with these LUX results.  We did not account for the energy resolution of the detector in this estimate, which would make this a conservative estimate for the number of events in the LUX exposure due to the rapid rise of the Foot model at small energies as discussed previously.

The XENON1T experiment also recently used an S2-only channel to search for ionization signals from light dark matter~\cite{Aprile:2019xxb}.  In this work the experiment found a rate of less than $\rm 1\, event/(tonne\times day\times keVee)$ above an energy of $\sim0.4\,$keVee in an exposure of 22~tonne-days.  This would correspond to roughly 130 events in the range 0.4--6~$\mathrm{keV}_{\mathrm{ee}}$  which includes all backgrounds in the detector.  Using the best fit Foot model to the DAMA data, Eq.~\ref{eq:countrate}  would predict approximately 2800 events in this exposure.   We have again not accounted for the resolution of the detector in this calculation.  We thus also find that this S2-only  analysis from XENON1T eliminates this model as a viable explanation for the DAMA annual modulation.

\section{Conclusion}
In this paper we have tested how well the plasma dark matter model proposed by Foot fits with the observed annual modulation signal in DAMA, by performing a goodness of fit test. In the model proposed by Foot, the dark matter halo surrounding galaxies consists of a plasma of dark electrons and protons charged under an unbroken $\rm U^{\prime}(1)$ symmetry. In 2014, Foot proposed that the DAMA modulation signal could be due to dark electrons in the Milky Way plasma recoiling off loosely bound electrons in the detector. 


Assuming these electron recoils gives a simple phenomenological model where the differential count rate is inversely proportional to the recoil energy, as in Eq.~\eqref{eq: diff recoil rate}.  Based on this model, we have performed a $\chi^2$ goodness-of-fit test with the observed modulation amplitudes in DAMA. We have found that the best-fit model is excluded at $\sim 7\sigma$ using this goodness of fit test. We also point out that both LUX and XENON1T have presented analyses of their data that are completely inconsistent with the predicted signal from the best fit Foot model for the DAMA data.  Thus, we conclude that the plasma dark matter model as proposed by Foot cannot explain the annual modulation observed by the DAMA collaboration.


\acknowledgments
SJ acknowledge support from the Swedish Research Council (Vetenskapsradet) through the Oskar Klein Centre (Contract No. 638-2013-8993).





\printbibliography

\end{document}